\documentclass[conference]{IEEEtran}
\IEEEoverridecommandlockouts
\usepackage{cite}
\usepackage{amsmath,amssymb,amsfonts}
\usepackage{subcaption} 
\usepackage{graphicx}
\usepackage{textcomp}
\usepackage{xcolor}
\usepackage{algorithm}
\usepackage{url}
\usepackage{tabularx} 
\usepackage{array}
\def\BibTeX{{\rm B\kern-.05em{\sc i\kern-.025em b}\kern-.08em
    T\kern-.1667em\lower.7ex\hbox{E}\kern-.125emX}}
\usepackage{algorithm}
\usepackage{algcompatible}
\begin{document}

\title{PhenoGnet: A Graph-Based Contrastive Learning Framework for Disease Similarity Prediction\\
}

\author{\IEEEauthorblockN{Ranga Baminiwatte}
\IEEEauthorblockA{\textit{School of Computing} \\
\textit{Clemson University}\\
\textit{Clemson, SC, U.S.A.}\\
}
\and
\IEEEauthorblockN{Kazi Jewel Rana}
\IEEEauthorblockA{\textit{School of Computing} \\
\textit{Clemson University}\\
\textit{Clemson, SC, U.S.A.}\\
}
\and
\IEEEauthorblockN{Aaron J. Masino}
\IEEEauthorblockA{\textit{School of Computing} \\
\textit{Clemson University}\\
\textit{Clemson, SC, U.S.A.}\\}
}

\maketitle

\begin{abstract}
Understanding disease similarity is critical for advancing diagnostics, drug discovery, and personalized treatment strategies. We present PhenoGnet, a novel graph-based contrastive learning framework designed to predict disease similarity by integrating gene functional interaction networks with the Human Phenotype Ontology (HPO). PhenoGnet comprises two key components: an intra-view model that separately encodes gene and phenotype graphs using Graph Convolutional Networks (GCNs) and Graph Attention Networks (GATs), and a cross-view model—implemented as a shared-weight multilayer perceptron (MLP)—that aligns gene and phenotype embeddings through contrastive learning. The model is trained using known gene–phenotype associations as positive pairs and randomly sampled unrelated pairs as negatives. Diseases are represented by the mean embeddings of their associated genes and/or phenotypes, and pairwise similarity is computed via cosine similarity. Evaluation on a curated benchmark of 1,100 similar and 866 dissimilar disease pairs demonstrates strong performance, with gene-based embeddings achieving an AUCPR of 0.9012 and AUROC of 0.8764, outperforming existing state-of-the-art methods. Notably, PhenoGnet captures latent biological relationships beyond direct overlap, offering a scalable and interpretable solution for disease similarity prediction. These results underscore its potential for enabling downstream applications in rare disease research and precision medicine.
\end{abstract}


\section{Introduction}
Our knowledge of Human diseases has evolved over the years, yet the molecular basis, phenotypic traits, and therapeutic targets of most diseases remain unclear. An increasing number of studies have observed that similar diseases often have related molecular causes, can be diagnosed by similar markers or phenotypes (set of observable traits that may be used to characterize a disease), or can be treated by similar drugs \cite{cheng2019}. Therefore, understanding similarity between diseases is a critical task in biomedical research, offering insights into shared pathological mechanisms and enabling advancements in disease diagnosis\cite{xiang2022}, drug discovery and repurposing\cite{wang2023}, and treatment planning \cite{bang2023}.

Current computational methods for assessing disease similarity can be categorized into three classes: (i) Molecule-based approaches, which leverage disease-associated molecular entities such as genes and proteins\cite{Suthram2010}, (ii) Semantic-based approaches, which use the hierarchical structure and term relationships in disease ontologies\cite{Resnik1995}, \cite{Lin1998}, \cite{Mathur2012}, (iii) Phenotype-based approaches, which utilize clinical phenotypic profiles and their qualitative associations with diseases\cite{Zhang2010},  \cite{Zhou2014}. While each of these approaches offers unique insights, they typically operate in isolation and therefore fail to capture complementary information from the others. For example, most molecule-based methods represent diseases solely at the genetic level, without taking into account both the semantic context provided by ontological hierarchies and the clinical relevance of phenotypic associations.

We introduce PhenoGnet, a deep learning framework that learns and refines latent semantic representations from ontological hierarchies and molecular level biological networks to support downstream tasks such as disease similarity estimation with a focus on rare diseases. Specifically, in this work, PhenoGnet utilizes graph neural networks to jointly integrate a gene functional interaction network and the Human Phenotype Ontology (HPO) \cite{Kohler2021} for accurate prediction of disease similarity. We first create 2 bio-entity networks including a functional gene interaction network, the HPO, and 3 association networks including gene-disease, HPO-disease and gene-HPO. Then we integrate them into a graph collaborative framework where the gene interaction network and the HPO graph are modeled as graph neural networks. Then we project the graph nodes into a common embedding space where contrastive learning is applied to maximize the embedding similarity between genes and HPO terms with known associations. The updated embeddings of the gene and HPO terms are used to predict disease similarity, where the diseases are represented as the average embeddings of their associated genes or HPO terms, and similarity was measured using cosine similarity between two disease representations. To fine-tune and validate this approach, we used 1,100 curated pairs of similar rare diseases and 866 randomly selected dissimilar pairs, with known gene and HPO associations. We found that this approach captures implicit disease information from known gene-phenotype associations, even without explicit representation in the model or training data.

\section{Related Work}

Graph Neural Networks (GNNs), as a specialized branch of deep learning designed for graph data, have demonstrated significant success, and are particularly well-suited for biomedical informatics. With the rapid growth of biological network data, GNNs have shown strong performance in tasks such as disease association prediction, medical image segmentation, and the identification of disease-gene relationships \cite{zhang2021}. However, the effectiveness of GNNs often relies on the availability of task centric labeled data, which can be scarce in some biomedical domains. To address this limitation, Graph Contrastive Learning (GCL) has emerged as a powerful approach for unsupervised or semi-supervised graph representation learning by allowing to learn discriminative embeddings without requiring extensive labels \cite{zhu2023}. 

Building on this, multi-view graph contrastive learning methods have been developed to further mitigate data sparsity and learn representations from multiple graphs by projecting them into a common latent space\cite{hassani2020}. Particularly in biological association prediction scenarios, integrating diverse networks of graph data can enhance model robustness and generalization \cite{kang2024}. Recent studies have demonstrated the utility of multi-view graph contrastive learning for disease similarity prediction. For instance, CoGO \cite{Chen2022} integrates Gene Ontology (GO) and gene interaction networks to generate disease representation embeddings to measure similarity. DisMVC \cite{Wei2024} combines gene interaction networks with miRNA similarity networks to improve similarity prediction. These approaches highlight the potential of leveraging multiple biological networks through contrastive objectives to capture implicit structural and semantic information.

\section{Methodology}\label{M:0}

\subsection{Data}\label{AA}

To capture the complex relationships between diseases, genes, and phenotypes, we construct five distinct association networks and integrate them into a unified graph-based collaborative framework.

The Human Phenotype Ontology (HPO) provides a standardized vocabulary of phenotype terms used to describe disease characteristics. The HPO graph ($G^p$) is modeled as a directed acyclic graph (DAG) that captures hierarchical relationships among phenotype terms.

The gene interaction network ($G^g$), sourced from HumanNet \cite{Kim2022}, is an undirected, weighted graph representing functional associations between genes. The edge weights reflect the statistical likelihood of gene-gene interactions.

Gene-phenotype and disease-phenotype associations were obtained from the HPO database, while disease-gene associations were derived from DisGeNET \cite{Pinero2017}. All three of these association networks are represented as undirected adjacency graphs. Table~\ref{tab:graphs} summarizes key statistics of the constructed graphs.

For training and evaluation, we use a curated benchmark dataset developed in prior studies on disease similarity \cite{Wei2024, dong2021, pakhomov2010}, \cite{Mathur2012}. It includes 1,100 rare disease pairs labeled as similar and 866 randomly selected dissimilar pairs, all with known gene and HPO associations. These are used for both fine-tuning and validation of our model.

The dataset is partitioned into training and validation sets as follows:

\begin{equation}
\begin{aligned}
\left\{
\begin{array}{l}
\mathcal{D}_{\text{train}} = \mathcal{D}_{\text{train}}^{\text{pos}} \cup \mathcal{D}_{\text{train}}^{\text{neg}} \\
\\
\mathcal{D}_{\text{validation}} = \mathcal{D}_{\text{validation}}^{\text{pos}} \cup \mathcal{D}_{\text{validation}}^{\text{neg}}
\end{array}
\right.
\end{aligned}
\end{equation}

\noindent such that $\mathcal{D}_{\text{train}} \cap \mathcal{D}_{\text{validation}} = \emptyset$. Specifically, $\mathcal{D}_{\text{train}}^{\text{pos}}$ and $\mathcal{D}_{\text{train}}^{\text{neg}}$ consist of 210 and 183 disease pairs, respectively, while $\mathcal{D}_{\text{validation}}^{\text{pos}}$ and $\mathcal{D}_{\text{validation}}^{\text{neg}}$ contain 890 and 683 pairs, respectively.

\begin{table}[H]
\caption{Graphs }
\label{tab:graphs}
\centering
\resizebox{\columnwidth}{!}{%
\begin{tabular}{|c|m{3.5cm}|c|c|}
\hline
\textbf{Graph} & \textbf{Description} & \textbf{No. of nodes} & \textbf{No. of edges} \\ \hline
$G^p$ & \centering Human phenotype ontology(HPO) & 19034 HPO terms  & 46784 \\ \hline
$G^g$ & \centering Gene interactions network & 18459 genes & 977495 \\ \hline
$G^{gp}$ & \centering Gene-phenotype associations & 18459 genes, 19034 HPO terms & 848284 \\ \hline
$G^{dp}$ & \centering Disease-phenotype associations & 30170 diseases, 19034 HPO terms & 171013 \\ \hline
$G^{dg}$ & \centering Disease-gene associations & 30170 diseases, 18459 genes & 807292 \\ \hline
\end{tabular}%
}
\end{table}

\subsection{Natural Language Processing of Term Descriptions}

To enhance the graph-based representation of the HPO, we construct a vector library $E^p \in \mathbb{R}^{n_p \times m}$,
where each row corresponds to an HPO term and \(m = 768\) denotes the embedding dimensionality. We generate these embeddings by encoding the textual descriptions of HPO terms (version 2025-01-16) using the Sentence-BERT framework~\cite{Reimers2019}, which maps text into continuous-valued vectors. Specifically, we employ the pre-trained \textit{all-mpnet-base-v2} model~\cite{Song2020}, accessed via Hugging Face at \url{https://huggingface.co/sentence-transformers/all-mpnet-base-v2}, without additional fine-tuning.

\subsection{Intra-view and Cross-view models}

\subsubsection{Network Construction}

For the HPO network \(G^p\), we construct a directed graph with unit-weight edges. We augment the graph using the true-path rule~\cite{Giorgio_2009}, connecting each term to all of its ancestor terms. This network is represented by the adjacency matrix \(\mathbf{A}^p \in \mathbb{R}^{n_P \times n_P}\), where \(n_P=19034\) is the number of terms in \(G^p\).

For the HumanNet gene functional interactions network \(G^g\), we build an undirected graph in which the edge weight \(w_{i,j}\) between genes \(g_i\) and \(g_j\) is defined as:
\begin{equation}
w_{i,j} = \frac{l_{i,j} - l_{\min}}{l_{\max} - l_{\min}}
\label{eq:network1}
\end{equation}
Here, \(l_{i,j}\) denotes the log-likelihood similarity from HumanNet, and \(l_{\min}\) and \(l_{\max}\) are the minimum and maximum log-likelihood similarities across all gene pairs. We represent this network by the adjacency matrix \(\mathbf{A}^g \in \mathbb{R}^{n_g \times n_g}\), where \(n_g=18459\) is the number of genes in \(G^g\).

\subsubsection{Intra-view Model}

Following previous work~\cite{Chen2021,Wei2024}, we model the gene interaction network \(G^g\) with a graph convolutional network (GCN) that learns node embeddings from both graph structure and node features. At the \((k+1)\)-th layer, the gene feature matrix \(\mathbf{f}^{g,k+1}\) is given by:
\begin{equation}
\mathbf{f}^{g,k+1} = \sigma\!\bigl(\mathbf{D}^{-\tfrac12}\,\hat{\mathbf{A}}^g\,\mathbf{D}^{-\tfrac12}\,\mathbf{f}^{g,k}\,\mathbf{W}^{k}\bigr)
\label{eq:network2}
\end{equation}
where \(\hat{\mathbf{A}}^g = \mathbf{A}^g + \mathbf{I}^g\) is the adjacency matrix of \(G^g\) with self-loops, and \(\mathbf{A}^g \in \mathbb{R}^{18{,}459\times18{,}459}\) contains the weight scores from~\eqref{eq:network1}. Here, \(\mathbf{D}\) is the diagonal degree matrix with entries \(D_{ii} = \sum_{j=1}^{n_g}A^g_{ij}\), \(\mathbf{W}^k\) is the learnable weight matrix at layer \(k\), and \(\sigma\) denotes the LeakyReLU activation function. We initialize \(\mathbf{f}^{g,0}\) as a one-hot encoding. After message-passing updates, the final matrix \(\mathbf{F}^G\) contains the latent representations of all genes in \(G^g\).

Similarly, we model the HPO network \(G^p\) as a graph attention network (GAT) that aggregates neighborhood information via attention coefficients~\cite{Wei2024}. At the \((k+1)\)-th layer, we compute the unnormalized coefficient between nodes \(i\) and \(j\) as:
\begin{equation}
e_{ij} = \sigma\!\bigl(\mathbf{a}^\mathsf{T}\,(\mathbf{W}^k\mathbf{f}_i^{p,k}\,\|\,\mathbf{W}^k\mathbf{f}_j^{p,k})\bigr)
\label{eq:network3}
\end{equation}
where \(\mathbf{a}\) is a learnable attention vector, \(\mathbf{W}^k\) is the layer-specific weight matrix, \(\sigma\) is LeakyReLU, and \(\mathbf{f}_i^{p,k},\mathbf{f}_j^{p,k}\) are the embeddings of HPO terms \(p_i\) and \(p_j\). We normalize these via softmax:
\begin{equation}
a_{ij} = \frac{\exp(e_{ij})}{\sum_{z \in \mathcal{N}_i}\exp(e_{iz})}
\label{eq:network4}
\end{equation}
with \(\mathcal{N}_i\) the first-order neighbors of \(p_i\). The updated embedding for node \(p_i\) then is
\begin{equation}
\mathbf{f}_i^{p,k+1} = \sigma\!\Bigl(\sum_{j \in \mathcal{N}_i}a_{ij}\,\mathbf{W}^k\mathbf{f}_j^{p,k}\Bigr)
\label{eq:network5}
\end{equation}
where the LeakyReLU activation \(\sigma\) follows the attention-weighted sum. This mechanism allows \(G^p\) to dynamically emphasize the most informative HPO term interactions during feature propagation. We set the initial feature matrix to contain the contain HPO sentence embeddings,  \(\mathbf{f}^{p,0}\) = $E^p$.After updates, the final matrix \(\mathbf{F}^P\) contains the latent representations of all phenotypes in \(G^p\).

\subsubsection{Cross-view model}
To align the gene and phenotype embeddings from our intra-view models, we apply a weight‐sharing nonlinear projection into a common latent space. Specifically, we pass the GAT and GCN embeddings through a single-hidden-layer multilayer perceptron (MLP) with shared weights. This projection is designed to extract joint features from gene–phenotype associations and to maximize agreement between genes and their corresponding ontology concepts.

We then train both graph models jointly using a cross-view contrastive loss function \cite{khosla2021}:
\begin{equation}
L^{\mathrm{cv}} \;=\; \beta\,L^{g} \;+\; (1-\beta)\,L^{p}
\label{eq:contrasiveloss1}
\end{equation}
where \(L^{g}\) and \(L^{p}\) represent the losses for the gene view and the phenotype view, respectively, and \(\beta\in[0,1]\) balances their contributions.

Following prior work \cite{Wei2024}, each view‐specific loss is computed as a normalized temperature‐scaled cross‐entropy over positive and negative pairs:
\begin{equation}
L^{g} \;=\; \sum_{i=1}^{n_g}
      -\log 
      \frac{\sum_{j=1}^{n_p} \mathbb{I}_{i,j}\,\exp\!\bigl[\operatorname{sim}(f_i^g, f_j^p)/\tau \bigr]}
           {\sum_{k=1}^{n_p} \exp\!\bigl[\operatorname{sim}(f_i^g, f_k^p)/\tau \bigr]}
\label{eq:contrasiveloss2}
\end{equation}
\begin{equation}
L^{p} \;=\; \sum_{i=1}^{n_p}
      -\log 
      \frac{\sum_{j=1}^{n_g} \mathbb{I}_{i,j}\,\exp\!\bigl[\operatorname{sim}(f_i^p, f_j^g)/\tau \bigr]}
           {\sum_{k=1}^{n_g} \exp\!\bigl[\operatorname{sim}(f_i^p, f_k^g)/\tau \bigr]}
\label{eq:contrasiveloss3}
\end{equation}
Here,
\[
\operatorname{sim}(x,y) = \frac{x \cdot y}{\|x\|\,\|y\|}
\]
is the cosine similarity; \(f_i^g\in\mathbb{R}^d\) and \(f_j^p\in\mathbb{R}^d\) are the projected feature vectors for the \(i\)-th gene and the \(j\)-th HPO term, respectively; \(\mathbb{I}_{i,j}=1\) if gene \(i\) and phenotype \(j\) are connected in the bipartite graph \(G^{gp}\) (and 0 otherwise); \(\tau\) is a temperature parameter controlling distribution sharpness; and \(n_g\), \(n_p\) are the numbers of genes and phenotypes.

During training, gene and HPO embeddings are continuously updated where positive pairs are pulled closer together in the latent space while negatives are pushed further from each other. Each true gene–phenotype pairing constitutes a positive example, while all other pairings serve as negatives. We optimize the total cross-view loss \(L^{\mathrm{cv}}\) using the RMSProp algorithm with an adaptive learning rate. At the end of training, the updated gene and HPO embeddings are preserved for the downstream task.
 
\begin{figure}[htbp] 
\centerline{\includegraphics[scale=0.5]{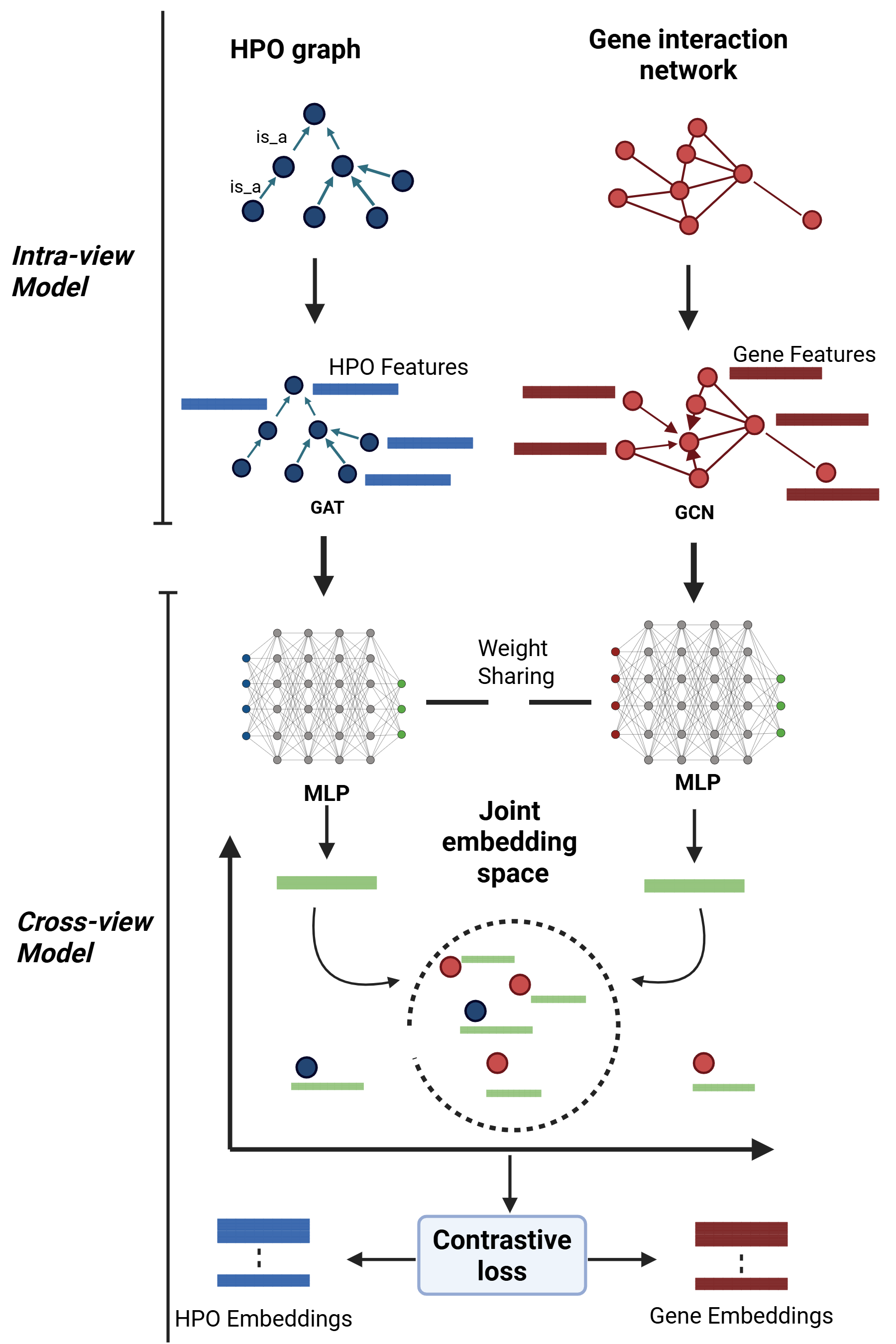}}
\caption{Overview of PhenoGnet:The intra-view model uses a GCN and a GAT for encoding gene and HPO terms respectively. Both networks have two graph layers (convolution or attention) and a leaky-ReLU activation. Genes and HPO terms are initially represented by one-hot encoding vectors which are transformed into a 32-dimensional embedding space after application of the GCN and GAT layers, respectively. The cross-view model is a shared-weight multilayer perceptron (MLP) model that aligns the gene and HPO latent spaces by projecting their respective embeddings into a shared latent space. Then, PhenoGnet is trained through contrastive-loss minimization, which maximizes the embedding similarity between genes and HPO terms with known associations. Finally, the trained HPO and gene embeddings are preserved to derive the disease representation vectors for the downstream task.}
\label{fig:gamma1-dist}
\end{figure}

\subsection{Disease pair similarity evaluation}

To evaluate pairwise disease similarity, we follow prior work and applied an average‐pooling strategy to derive a single feature vector per disease.  For each disease \(d_z\), let \(\mathbb{G}_z\) denote its associated genes (from \(G^{dg}\)) and \(\mathbb{P}_z\) denote its associated HPO terms (from \(G^{dp}\)).  We then computed the gene‐based and phenotype‐based disease embeddings as
\begin{equation}
f_{d_z}^g
=\frac{1}{\lvert \mathbb{G}_z\rvert}
\sum_{i\in\mathbb{G}_z} f_{i}^g,
\label{eq:dissim1}
\end{equation}
\begin{equation}
f_{d_z}^p
=\frac{1}{\lvert \mathbb{P}_z\rvert}
\sum_{j\in\mathbb{P}_z} f_{j}^p,
\label{eq:dissim2}
\end{equation}
where \(f_{i}^g\in F^G\) and \(f_{j}^p\in F^P\) are the learned feature representations for gene \(i\) and phenotype term \(j\), respectively.  Additionally  we concatenate these two views into a combined disease embedding,
\begin{equation}
f_{d_z}
=\mathrm{concat}\bigl(\,\gamma\,f^g_{d_z},\;(1-\gamma)\,f^p_{d_z}\bigr),
\label{eq:dissim3}
\end{equation}
in which the scalar \(\gamma\in[0,1]\) governs the relative weighting of the gene‐based vector \(f^g_{d_z}\) (Eq.~\ref{eq:dissim1}) versus the phenotype‐based vector \(f^p_{d_z}\) (Eq.~\ref{eq:dissim2}).  

Finally, for any two diseases \(d_i\) and \(d_j\), we compute their similarity as the cosine of the angle between their combined embeddings:
\begin{equation}
\mathrm{sim}(d_i,d_j)
=\frac{f_{d_i}\,f_{d_j}^T}{\|f_{d_i}\|\;\|f_{d_j}\|}.
\label{eq:dissim4}
\end{equation}

\subsection{Hyper-parameter optimization}

We utilize Optuna \cite{optuna_2019}, a Python-based framework for automated hyperparameter optimization, to determine the optimal values for several key hyperparameters, including the contrastive‐loss weight \(\beta\), the gene–phenotype mixing coefficient \(\gamma\), the temperature parameter \(\tau\), the learning rate, and the number of training epochs for the RMSProp optimizer. To ensure robustness, we adopt a 5-fold cross-validation strategy on the training dataset $\mathcal{D}_{\text{train}}$. For each fold, the model was trained on 80\% of the disease pairs and validated on the remaining 20\%. A total of 30 Optuna trials were executed, with each trial corresponding to a unique combination of hyperparameters sampled via the Tree-structured Parzen Estimator (TPE) algorithm. Model performance in each trial is evaluated using AUROC and AUPRC metrics, and the final objective combined both via a harmonic mean to reflect balanced classification performance. The best hyperparameter configuration is selected based on the trial that maximized this combined metric across all folds.

\section{Results and Discussion}

\subsection{Disease similarity prediction}

We used three types of disease representation embeddings to calculate disease similarity on $\mathcal{D}_{\text{validation}}$: gene-based embeddings $(f^g_{d_z})$, phenotype-based embeddings $(f^p_{d_z})$, and their combination $(f_{d_z})$. Furthermore, to measure the effectiveness of these embeddings, we compare the performance with a replication of CoGO \cite{Chen2022}, an existing state-of-the-art (SOTA) method for measuring disease similarity. Table \ref{tab:results} shows a summary of the validation results.

\begin{figure}[htbp]
    \centering
    \begin{subfigure}[t]{0.44\textwidth}
        \centering
        \caption{Precision-recall curves}
        \includegraphics[width=\linewidth]{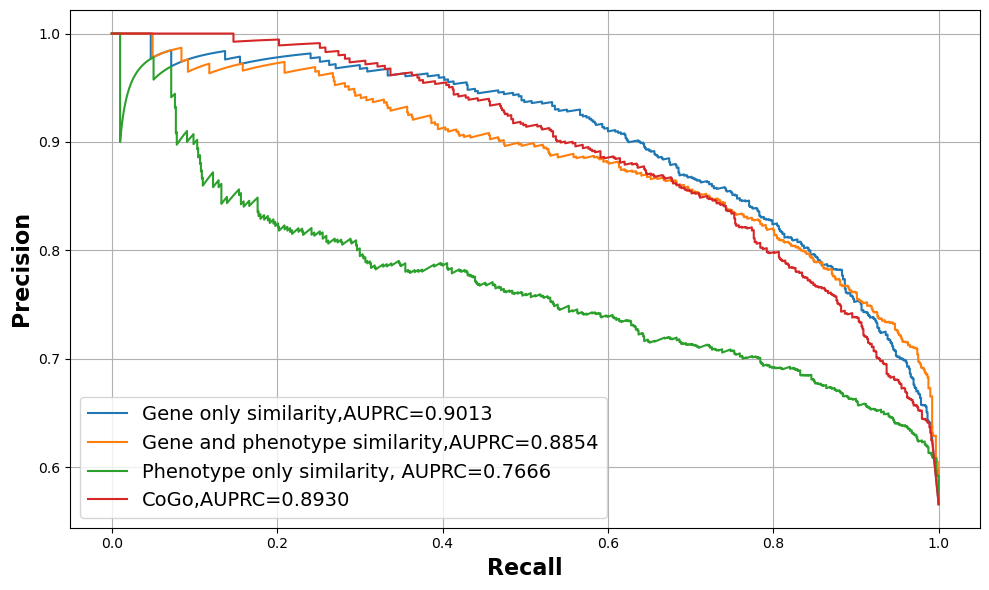}
        \label{fig:plot1}
    \end{subfigure}
    \hfill
    \begin{subfigure}[t]{0.44\textwidth}
        \centering
        \caption{ROC-curves}
        \includegraphics[width=\linewidth]{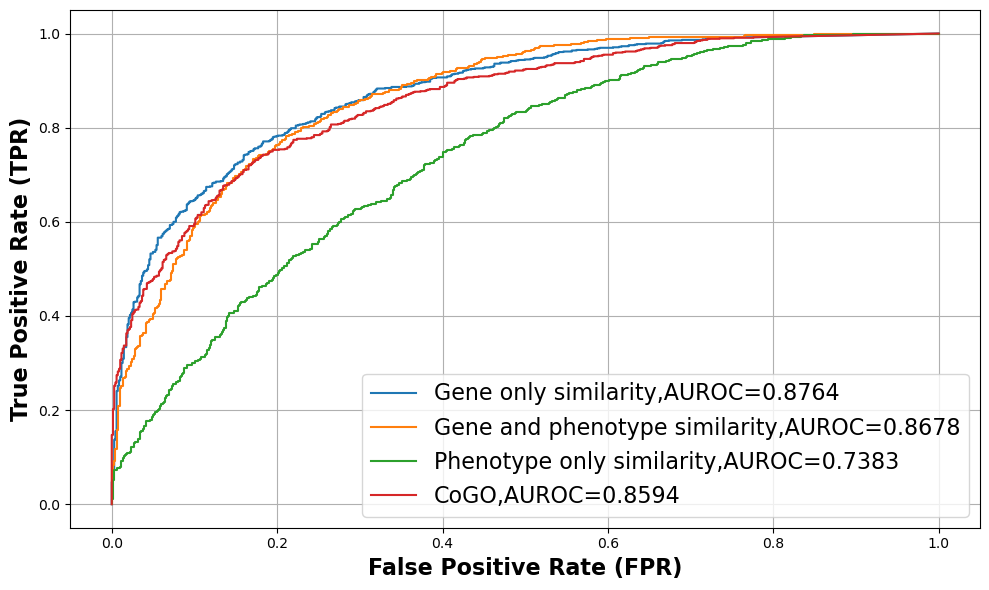}
        \label{fig:plot2}
    \end{subfigure}
    
    \caption{Precision-recall curves and ROC-curves for types of disease embeddings and previous SOTA method.}
    \label{fig:two-side-by-side}
\end{figure}

In terms of AUROC (Area Under the Receiver Operating Characteristic curve) and AUPRC (Area Under the Precision-Recall Curve), our framework closely matches the performance of CoGO (Fig. \ref{fig:two-side-by-side}). The gene-based embeddings achieve the highest AUPRC (0.9012) and AUROC (0.8764), outperforming both the combined (AUPRC = 0.8855, AUROC = 0.8678) and HPO-based embeddings (AUPRC = 0.7668, AUROC = 0.7383), as well as CoGO (AUPRC = 0.8928, AUROC = 0.8594). The F1 score is also highest for gene-based embeddings (0.8254), followed by the combined embeddings (0.8229) and CoGO (0.7983), indicating a well balanced performance.

Interestingly, while HPO-based embeddings achieve the highest sensitivity (0.9393), they show the lowest specificity (0.3353), indicating a trade-off between detecting true positives and avoiding false positives. In contrast, gene-based embeddings provide a more balanced sensitivity (0.8843) and specificity (0.6633), leading to a more robust overall performance. These results suggest that gene-based features contribute the most discriminative power in our similarity measurement framework, while HPO-based features capture broader phenotype-level associations. 

\begin{table}[H]
\caption{Validation results table}
\label{tab:results}
\centering
\renewcommand{\arraystretch}{1.3}
\resizebox{\columnwidth}{!}{%
\begin{tabular}{c | c c c c c}
\textbf{Method} & \textbf{AUPRC} & \textbf{AUROC} & \textbf{F1 score} & \textbf{Sensitivity} & \textbf{Specificity} \\
\hline
Gene based embeddings & 0.9012 & 0.8764 & 0.8254 & 0.8843 & 0.6633 \\
Combined embeddings  & 0.8855 & 0.8678 & 0.8229 & 0.9112 & 0.6047\\
HPO based embeddings & 0.7668 & 0.7383 & 0.7670 & 0.9393 & 0.3353 \\
CoGO & 0.8928 & 0.8594 & 0.7983 & 0.9247 & 0.4890\\
\end{tabular}%
}
\end{table}

\subsection{Similarity scores distribution}
\begin{figure}[htbp]
    \centering
    \begin{subfigure}[t]{0.24\textwidth}
        \centering
        \includegraphics[width=\linewidth]{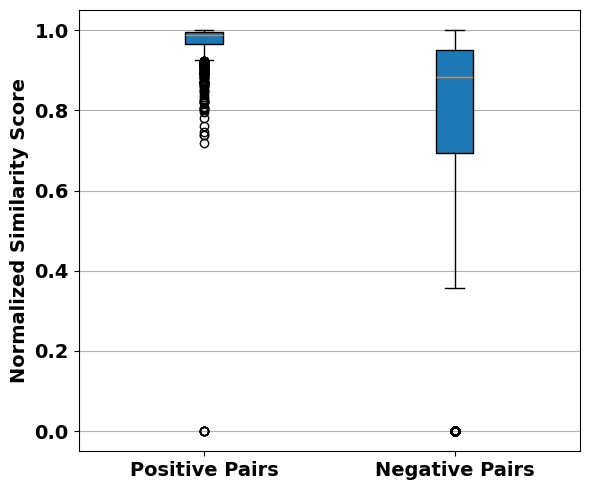}
        \caption{Gene embeddings}
        \label{fig:dist1}
    \end{subfigure}
    \hfill
    \begin{subfigure}[t]{0.24\textwidth}
        \centering
        \includegraphics[width=\linewidth]{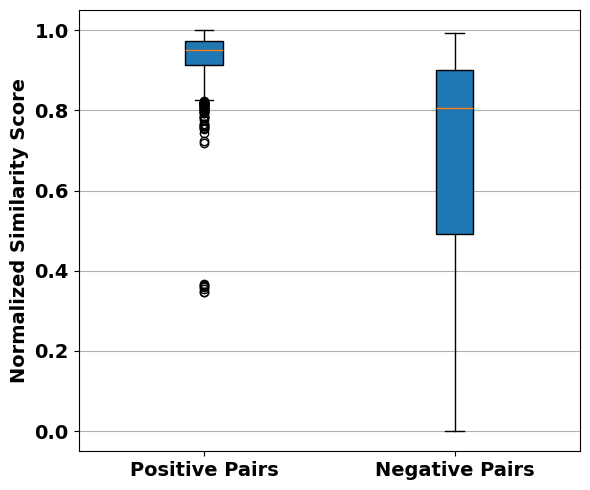}
        \caption{Combined embeddings}
        \label{fig:dist2}
    \end{subfigure}
    \hfill
    \begin{subfigure}[t]{0.24\textwidth}
        \centering
        \includegraphics[width=\linewidth]{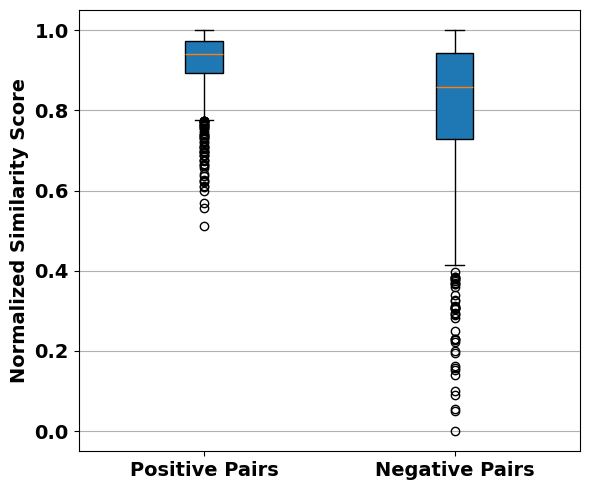}
        \caption{HPO embeddings}
        \label{fig:dist3}
    \end{subfigure}
    
    \caption{Distribution of similarity scores on similar and dissimilar pairs.}
    \label{fig:distributions}
\end{figure}

\begin{figure}[htbp]
    \centering
    \includegraphics[width=0.90\linewidth]{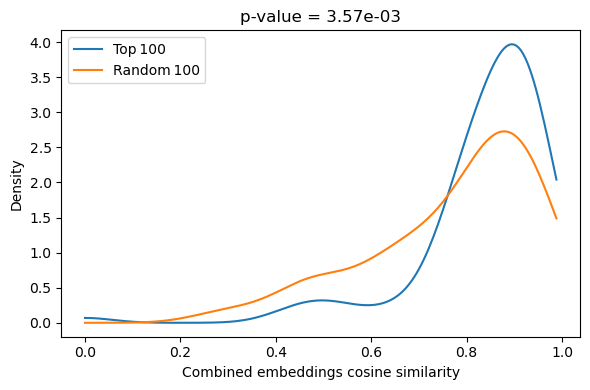}
    \caption{Distribution of maximum
similarity scores for top 100 similar pairs against 100 randomly
selected disease pairs.}
    \label{fig:top100vsrandom100}
\end{figure}

We plotted the the distribution of similarity scores predicted on disease pairs using all three types of disease representations(Fig. \ref{fig:distributions}). Similar (Positive) disease pairs achieve a significantly higher similarity scores across all three representations. One possible reasoning for high similarity score among similar disease pairs could be the direct overlap of associated gene and HPO terms between two diseases. Therefore, to assess direct overlap, we examined the top 100 most similar disease pairs and found that, on average, only 11.41\% of genes and 7.19\% of HPO terms were shared among disease representations, suggesting that our model captures relationships beyond explicit term sharing. We then removed all overlapping gene and HPO terms, and compared the distribution of maximum similarity scores for these 100 pairs against 100 randomly selected disease pairs; the mean maximum score for the top 100 remained significantly higher than that of the random set (Fig. \ref{fig:top100vsrandom100}). Together, these findings indicate that PhenoGnet infers latent gene–phenotype associations even in the absence of direct overlap. 

\subsection{Future work}

We plan to extend our framework to incorporate diverse biomedical knowledge sources, including protein-protein interaction networks, gene expression networks, and additional biological ontologies. This integration aims to enhance the biological contextualization of embeddings and support more detailed disease similarity assessments. Furthermore, to improve our model's understanding of complex graph structures, we aim to analyze bottleneck effects in the phenotype and gene networks. Techniques such as shortest path-based attention mechanisms or structural bottleneck scoring will be explored to identify critical nodes and edges that contribute most to disease similarity. We also plan to apply our framework to large-scale biobank datasets (e.g., UK Biobank) to identify representative and potentially novel phenotype sets associated with complex diseases. Case studies will be conducted to validate the biological and clinical relevance of these latent phenotype associations.

\section{Conclusion}

In this work, we introduced PhenoGnet, a graph-based contrastive learning framework designed to measure disease similarity by integrating gene interaction networks and the HPO. By leveraging intra-view GCN and GAT encoders and aligning gene and phenotype embeddings through a cross-view contrastive learning technique, our model effectively captures both explicit and implicit biological associations.

Experimental results demonstrate that PhenoGnet not only achieves performance comparable to, and in some cases exceeding, the current state-of-the-art methods, but also reveals disease relationships that extend beyond direct gene or phenotype overlaps. This suggests that PhenoGnet is capable of learning latent biological patterns and embedding detailed disease characteristics.

Our findings highlights the value of joint gene–phenotype modeling in supporting scalable and interpretable disease similarity analysis. As a result, PhenoGnet holds promise for advancing rare disease research, disease diagnosis, and precision medicine applications.

\section*{Acknowledgment}

We thank the NIH funded Center of Biomedical Research Excellence in Human Genetics at Clemson University for funding support.

\bibliographystyle{IEEEtran}
\bibliography{IEEEabrv,icmla-2025}
\end{document}